\documentclass[sigconf]{acmart}

\AtBeginDocument{%
  \providecommand\BibTeX{{%
    \normalfont B\kern-0.5em{\scshape i\kern-0.25em b}\kern-0.8em\TeX}}}

\usepackage{amsfonts}
\usepackage{amsmath}
\usepackage{hyperref}
\hypersetup{colorlinks=false,breaklinks=true}
\usepackage{subfigure}
\usepackage{subfloat}
\usepackage{graphicx}
\usepackage{textcomp}
\usepackage{xargs}                      
\usepackage{datetime}
\usepackage{float}
 
\usepackage{dsfont}
\usepackage{epsfig}
\usepackage{epstopdf}
\usepackage{multirow}
\usepackage{tabularx}
\usepackage{comment}
\usepackage[normalem]{ulem}

\usepackage{algorithm}
\usepackage[]{algpseudocode}
\usepackage{algorithmicx}
\algnewcommand\algorithmicforeach{\textbf{for each}}
\algdef{S}[FOR]{ForEach}[1]{\algorithmicforeach\ #1\ \algorithmicdo}
\algtext*{EndWhile}
\algtext*{EndIf}
\algtext*{EndFor}

\graphicspath{{./Figures/}}

\setcopyright{none}
\renewcommand\footnotetextcopyrightpermission[1]{} 

\begin{document}

\title{Learning Assisted Side Channel Delay Test for Detection of Recycled ICs}
\author{Ashkan Vakil}
\affiliation{\institution{George Mason University}}
\email{avakil@gmu.edu}

\author{Farzad Niknia}
\affiliation{\institution{\hspace{-0.5cm}University of Maryland Baltimore County}}
\email{farzadn1@umbc.edu}

\author{Ali Mirzaeian}
\affiliation{\institution{George Mason University}}
\email{amirzaei@gmu.edu}


\author{Avesta Sasan}
\affiliation{\institution{George Mason University}}
\email{asasan@gmu.edu}

\author{Naghmeh Karimi}
\affiliation{\institution{\hspace{-0.5cm}University of Maryland Baltimore County}}
\email{nkarimi@umbc.edu}

\renewcommand{\shorttitle}{ }

\begin{abstract}\label{sec:abs}
With the outsourcing of design flow, ensuring the security and trustworthiness of integrated circuits has become more challenging. Among the security threats, IC counterfeiting and recycled ICs have received a lot of attention due to their inferior quality, and in turn, their negative impact on the reliability and security of the underlying devices. Detecting recycled ICs is challenging due to the effect of process variations and process drift occurring during the chip fabrication. Moreover, relying on a golden chip as a basis for comparison is not always feasible. Accordingly, this paper presents a recycled IC detection scheme based on delay side-channel testing. The proposed method relies on the features extracted during the design flow and the sample delays extracted from the target chip to build a Neural Network model using which the target chip can be truly identified as new or recycled. The proposed method classifies the timing paths of the target chip into two groups based on their vulnerability to aging using the information collected from the design and detects the recycled ICs based on the deviation of the delay of these two sets from each other. 
\end{abstract}

\pagestyle{plain} 

\keywords{Counterfeit IC, Aging, Recycled IC, Hardware Security, Learning}
\settopmatter{printacmref=false}

\maketitle

\section{Introduction}\label{sec:Intro} 
To reduce the fabrication cost of a new Integrated Circuit (IC), benefit from smaller area and power profile of advanced sub-nanometer technology nodes, or build a scalable model that is ready to meet the market demand, the manufacturing supply chain of ICs is globalized \cite{yeh2012trends}. However, the presence of untrusted parties in such a global supply chain has jeopardized the security and trustworthiness of Integrated Circuits (ICs) and introduced many security vulnerabilities \cite{threats}, including the possibility of IP piracy/theft, Trojan insertion, overbuilding, and counterfeiting. A widely studied solution for protecting the IP in the manufacturing supply chain is logic encryption. For logic encryption, the IC designer introduces limited means of post-manufacturing programmability into a netlist (representing valuable IP) \cite{lutlock,dfssd,scramble,srclock,fullLock,interlock,luticcad}. After fabrication, the IC is shipped to and activated at a trusted facility, where the activation key is stored in a non-volatile memory on-chip. Logic locking protects an IC against overproduction and IP piracy and reduces Trojan insertion chances, but does not protect recycled or aged ICs sold as new.   

A counterfeit IC can be an unauthorized copy of the exact IC, a slightly modified version, a recycled IC, or a defective one \cite{guin2014counterfeit}. 
Recycled ICs, ICs that have been already used but are pretended to be new, have contributed to more than 80\% of the counterfeit ICs in recent years\cite{guin-13-selection}; posing around \$169 billion revenue loss to the global electronics supply chain\cite{alam_18_robust}. Based on U.S. Chamber of Commerce reports, counterfeit ICs have even found their way into military supply chain~\cite{uscc}. A recycled chip exhibits a lower performance and a higher failure rate over time, since its embedded transistors have already been aged, i.e., their characteristics have been derated due to the chip usage. This increases the probability of chip failure sooner than expected; shortening the chip lifespan. The short lifetime and low performance of the recycled ICs not only affect the end-users but also puts a significant financial burden on the industry and government sectors. Therefore, a concrete solution to detect recycled ICs is highly required. 

Among the various designs' robustness concerns affecting CMOS devices, aging effects have been receiving a lot of attention. In practice, aging mechanisms degrade the reliability and performance of CMOS devices over their lifetime. Due to aging, the electrical behavior of transistors deviates from their original intended behavior, resulting in performance degradation and the ultimate chip failure~\cite{karimi2018effect}. Among aging mechanisms, the effect of Bias Temperature Instability (BTI) and Hot Carrier Injection (HCI) are more dominant than other aging mechanisms~\cite{oboril-12-extratime}. In this paper, we focus on these two aging mechanisms when detecting the recycled chips. 

\begin{figure}[h]
  \centering
   \includegraphics[width=2.6552in]{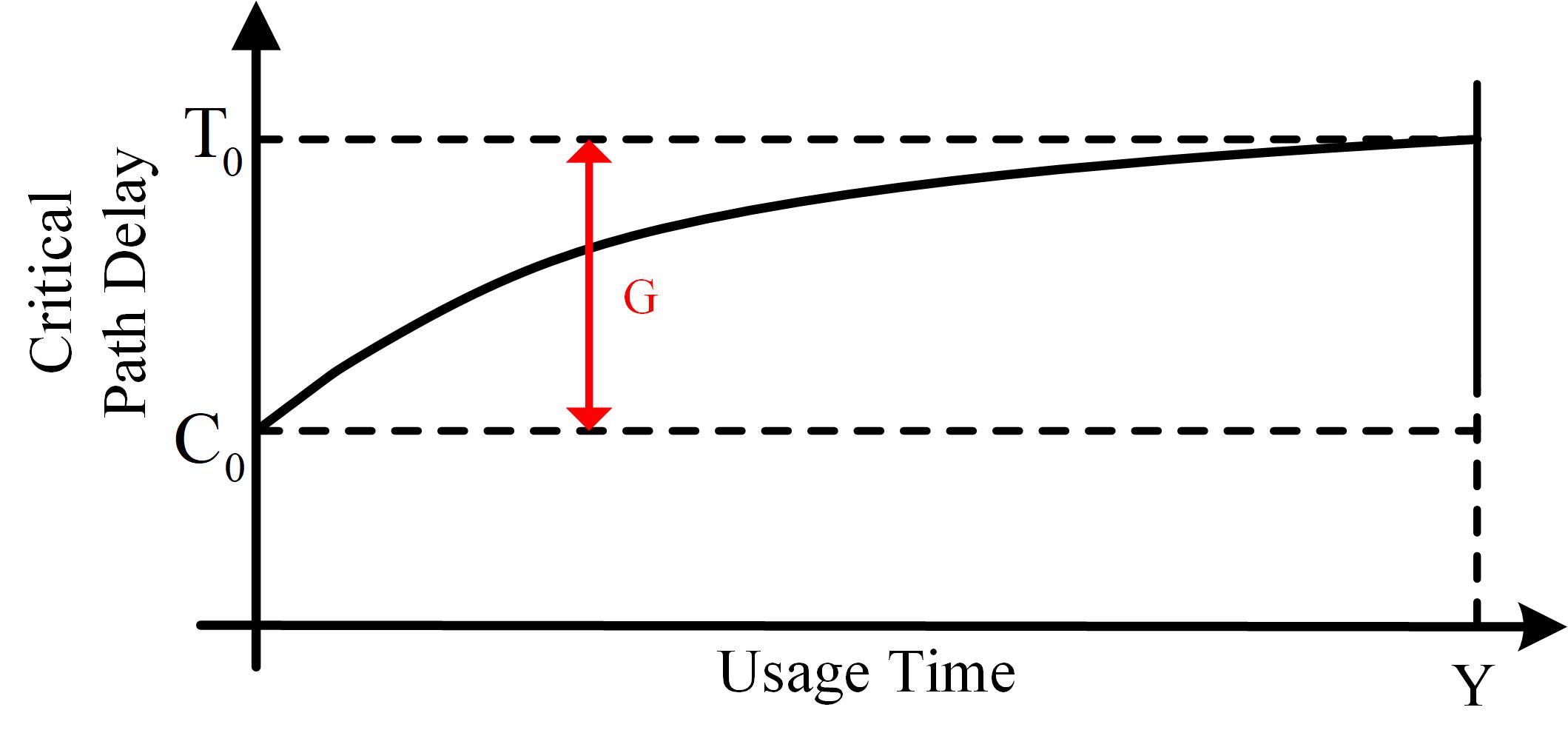}
   \caption{Impact of aging on critical path delay.}
   \vspace{-12pt}
    \label{fig:magic}
\end{figure}

Guardbanding is the current industrial practice to cope with transistor aging and voltage droops. It entails slowing down the clock frequency (i.e., adding timing margin during design) based on the worst degradation the transistors might experience during their lifetime~\cite{karimi-15-magic}. The guardbands ensure that the chip functionality is intact for an average period of 5 to 7 years. However, inserting wide guardbands degrades performance and increases energy consumption. Hence, chip design companies usually have small guardbands, typically 5-10\%~\cite{ASPDAC}. Fig.~\ref{fig:magic} shows an overview of the guardband assignment for each chip during the manufacturing. In this figure, the delay of the critical path is $C_0$ when the chip is new. As the chip is used, its critical path delay gradually increases due to aging; reaching $T_0$ after a period of t = Y years. Accordingly, for the chip to be usable during its expected lifetime (Y Year), it needs to be clocked at a frequency no higher than $F_0$ = 1/$T_0$. Thus, the designers add a guardband $G = T_0-C_0$ to prevent any aging-induced chip malfunction during its expected lifespan (i..e, Y years in Fig.~\ref{fig:magic}). However, if the chip is a recycled one, the remaining life expectancy is less than Y, i.e., it can experience a malfunction much sooner than the expected lifetime (Y).

In practice, detecting recycled ICs is challenging, particularly in lower technology nodes, due to the Process Variations (PV) occurring at the fabrication process. To account for the PV, the detection method should be able to differentiate the PV-induced change in the delay of timing-paths from the aging-induced changes. Otherwise, ignoring the impact of PV would result in false positives or false negatives, which in turn results in disposing a new IC or deploying a recycled one, respectively.

Device workload as well as environmental conditions, including temperature and voltage source, all affect the device aging rate. In fact, each logical cell residing in a chip is aged differently based on the duration of logical `1' and `0' values applied to its input pins, or the toggling count that its transistors observe.

Our Learning Assisted Side Channel Delay Test scheme aims to differentiate between aged and new devices by assessing the impact of aging on the delay of carefully selected timing paths in a chip under test. Different timing paths, for having different topologies, for being made of different types of devices, and for experiencing different switching activities would age differently. Hence, in an aged IC, one could expect to see a disparity in the aging-induced increase in the delay across timing-paths. Therefore, delay testing could be used for the detection of aged ICs. 

The problem of finding aged ICs using delay testing, however, faces a serious challenge: The change in the delay of timing paths could also be the result of PV (systematic or random) and/or process drift. PV is the \textit{unintended} variation in the physical and electrical property (strength) of transistor devices due to the physical limitation of manufacturing devices in building perfect transistors. Process drift is the \textit{intended} change in the process over time, made by the fabrication lab to improve the process. Although such improvement guarantees a working transistor made using Spice models generated for an older version of the process, the strength and characteristic of devices (and in the result their speed) would change over time. Therefore, by having access to the original GDSII and timing model (generated for the original Spice Model), a tester can not determine if the change in the delay of a timing path is due to PV and/or process drift or if it is due to aging. 

In this paper, we propose a learning assisted mechanism for side-channel delay testing of a circuit to assess if the chip under test is aged. Our proposed solution is resilient to the PV and process drift and does not need a golden chip to work properly.

\section{Threat Model}\label{sec:threat}
In this paper, we assume that the IC designer is trustworthy while the supply chain is untrusted. In particular, we assume that an adversarial supplier can potentially provide the system integrator with a recycled IC, the usage of which (in a critical application) may result in catastrophic consequences (due to the effect of aging on the chip's reliability). We assume that the IC/system designer has access to the chip netlist and its GDSII file. However, she does not have a golden chip to use it as a basis to determine if the chip-under-test is recycled or new.

\section{Previous Works}\label{sec:Prev} 

Recycled IC detection methods can be classified into several categories. The first category deals with conventional test methods that perform physical (e.g. detecting repackaged ICs using 3-Dimensional imaging technologies) and electrical (studying the ICs' parameters such as threshold voltage, path delays, etc) tests~\cite{tehranipoor2015counterfeit}~\cite{shakya2015performance}. Such tests are conducted in specific labs following several testing standards such as AS6171, AS5553, and CCAP-101~\cite{SAE}. These methods are costly, time-consuming, and have a low detection rate.

The second category of Recycled IC detection schemes, i.e., statistical approaches, mainly deploy machine-learning models to differentiate new and aged chips from each other. For instance, in ~\cite{huang-15-recycled}, Ke Huang et al. used a Support Vector Machine (SVM) based technique to classify the chips into recycled versus new using parametric measurements collected from a set of brand new chips including Iddq, $F_{max}$, $V_{min}$, etc. These measurements are usually collected in the test facilities to verify the correct functionality of the chips before shipping them to customers. On the other hand, in~\cite{zheng-14-caci}, the authors detect recycled ICs based on the aging rates in similar components that may have resided in the target chip. The assumption is that if the device is aged, similar components in the device (e.g., different full adders in an N-bit adder module) may age differently as they are exposed to different input patterns during run time. In this method, the correlation of dynamic supply current (\textit{IDDT}) between similar logic blocks is calculated, based on which, the IC is reported as new or recycled. The main drawback of the prior art statistical solutions is their reliance on the availability of a golden chip.

The third group, the so-called DFAC (Design For Anti Counterfeiting) strategies, detect the recycled ICs via on-chip sensors\cite{shakya2015performance}. The on-chip sensor-based approaches try to compare the frequency of an embedded element, e.g., a Ring Oscillator (RO) with a reference point to identify the recycled ICs. For instance, Guin et al. \cite{guin2015design} proposed to insert two ring oscillators inside the chip, one is not used frequently while the other is always on. Comparing the frequency of these oscillators reveals if the chip is recycled. As this is sensitive to PV, its detection accuracy is low. To resolve this issue, ~\cite{alam_18_robust} tried to mitigate the impact of PV by replacing the reference RO with a Non Volatile Memory (NVM) that stores the frequency of the RO when it is new. The stored value is compared with the frequency of the RO when the device is checked regarding its freshness. Any discrepancy demonstrates that the device is not new. To prevent tampering such NVM, the authors proposed to use digital signature verification (e.g., chip unique ID). This method suffers from high power consumption related to its always-on RO.


\section{Preliminary Backgrounds}\label{sec:Prelim} 
\subsection{Background on aging.}
Aging mechanisms including Bias Temperature Instability (BTI), Hot-Carrier Injection (HCI), Time-Dependent Dielectric Breakdown (TDDB), and Electromigration (EM) result in performance degradation and eventual failure of digital circuits over time~\cite{kim-15-chip}. Among all, BTI and HCI are the two leading factors in the performance degradation of digital circuits~\cite{oboril-12-extratime}. Both mechanisms increase the switching delay of transistors, leading to an increase in path delays.

\textbf{BTI Aging:} BTI aging includes Negative Bias Temperature Instability (NBTI) and Positive Bias Temperature Instability (PBTI). 
NBTI affects a PMOS transistor when a negative voltage is applied to its gate. A PMOS transistor experiences two phases of NBTI depending on its operating condition. The first phase, the so-called \textit{stress phase}, occurs when the transistor is on ($V_{gs} < V_t$). In this case, positive interface traps are generated at the Si-SiO$_2$ interface which leads to an increase of the threshold voltage of the transistor. The second phase, denoted as the \textit{recovery phase}, occurs when the transistor is off ($V_{gs} > V_t$). The threshold voltage drift that occurred during the stress phase will partially recover in the recovery phase.

\begin{figure}[h]
  \centering
   \includegraphics[width=2.6552in]{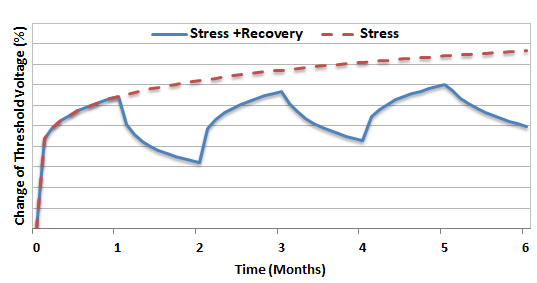}
   \vspace{-1em}
   \caption[Threshold-voltage shift of a PMOS transistor under NBTI effect]{Threshold-voltage shift of a PMOS transistor under the NBTI effect. Values on the Y-axis are not shown to make the graph generic for different technologies. \vspace{-5mm}}
    \label{fig:stress_recovery}
\end{figure}

Threshold voltage drifts of a PMOS transistor under stress depend on the physical parameters of the transistor, supply voltage, temperature, and stress time~\cite{khan-11-nbti}. The last three parameters (known as external parameters) are used as acceleration factors of the aging process. Figure~\ref{fig:stress_recovery} shows the threshold voltage drift of a PMOS transistor that is continuously under stress for 6 months and a transistor that alternates between stress and recovery phases every other month. As shown, the NBTI effect is high in the first couple of months but the threshold voltage tends to saturate for long stress times. It is noteworthy to mention that PBTI affects the NMOS transistors in a similar fashion that NBTI affects the PMOS transistors. Accordingly, for the sake of space, we do not discuss PBTI separately. 
\vspace{-6mm}

\textbf{HCI Aging:} HCI occurs when hot carriers are injected into the gate dielectric during transistor switching and remain there. HCI is a function of switching activity and degrades the circuit by shifting the threshold voltage and the drain current of transistors under stress.
HCI mainly affects NMOS transistors.

HCI-induced threshold voltage drift is highly sensitive to the number of transitions occurring in the gate input of the transistor under stress.
In practice, HCI has a sublinear dependency on the clock frequency, usage time, and activity factor of the transistor under stress, where activity factor represents the ratio of the cycles the transistor is switching and the total number of cycles the device is utilized. HCI effect is exacerbated as the operating temperature increases~\cite{oboril-12-extratime}.

\section{Proposed Recycled IC Detection Methodology}\label{sec:proposed} 

Our proposed aged-IC detection methodology is based on a side-channel delay analysis. To formulate a reliable test, we have provided separate solutions for mitigating the impact of process drift and process variation, and have proposed a novel (delay-based frequency sweeping) test solution to detect the aged-ICs. For the use of machine learning models, our proposed solution falls in the category of statistical models. However, unlike prior work, we do not rely on the existence of a golden model. We, instead, assume that the designer has access to the original physical design, the GDSII sent for fabrication, and Static Timing Model developed for the original design at the state of its timing-closure. In the following subsection, we define a few terms and parameters, based on which in Section~\ref{proposed_model_section}, our proposed model is explained:
\vspace{-2mm}

\subsection{Definitions and Model Parameters}
 
Before describing our solution, we elaborate on the model parameters used:

\textbf{Clock Frequency Sweeping Test (CFST):} An existing delay testing solution in which delay of different timing paths is examined while increasing the clock frequency~\cite{cfst}. The target is to find the start to fail frequency for different timing paths. The test accuracy is limited by the tester frequency step size and maximum achievable frequency. The delay reported for each timing path may be affected by both process variation and process drift.  

\textbf{Age Distinguished Paths (ADP)}:
Depending on the circuit topology and workload, some of the timing paths in a circuit age more, and some age less than others. Hence, we can distinguish between two sets of timing paths: 1) Most aging Affected Paths (MAP) and 2) Least aging Affected Paths (LAP). 

\begin{figure}[h]
\centering
  \includegraphics[width=\columnwidth]{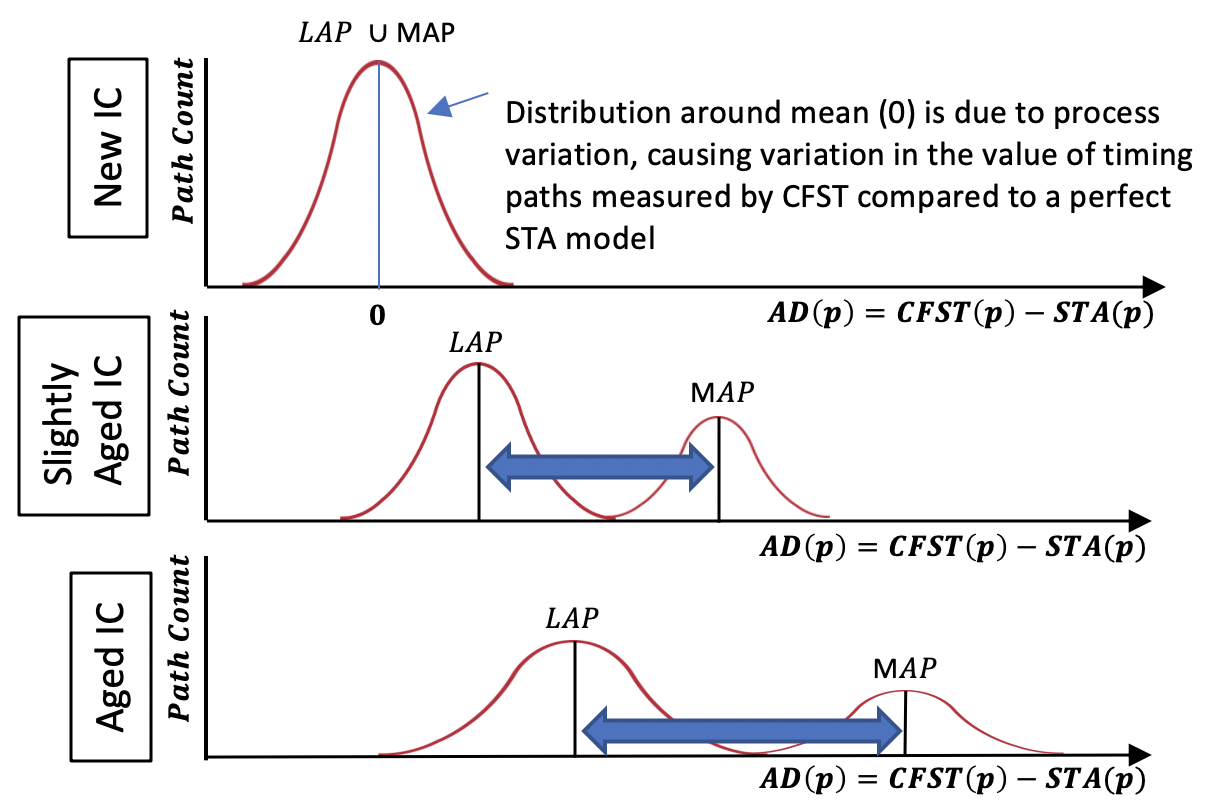}
  \caption{In a new device, one cannot distinguish between MAP and LAP timing paths as no aging occurred. Computing the $AD$ for timing paths gives us a zero-mean distribution. As the IC ages, the delay of all timing paths increases, however, the delay-increase is more significant in the MAP set of timing paths. Therefore, the normal distribution of $AD$ morphs into a bimodal distribution as the IC ages. Identification of MAP and LAP sets of timing paths allows us to compute the mean for each set. The shift in the mean is an indication of the extent of aging. In this figure, it is assumed that there is no process drift (but there exist process variation), the step size of the tester is very small, STA is perfect, and CFST reported delay for a timing path at age zero (fresh IC) matches STA. We will update these assumptions to realistic ones when discussing our proposed solution. \vspace{-4mm}}
 \label{fig:histogram_differences}
\end{figure}

For simplicity, lets first assume that there is no process drift (but there exist process variation), the step size of the tester is sufficiently small, Static Timing Analysis is perfect, and CFST reported delay for a timing path at age zero (fresh IC) matches that of the STA within the boundary of process variation. Lets denote the STA-reported delay of path $p$ with $STA(p)$, and the delay reported by CFST by $CFST(p)$. We define added delay $AD$ for path $p$ as $AD(p) = CFST(p)-STA(p)$. Given a set of timing paths, if we compute $AD$ for each path, we will see a zero-mean normal distribution of $AD$s if the IC is not aged (in an ideal world). This is illustrated in Fig. ~\ref{fig:histogram_differences}(top). As the IC ages, the MAP and LAP timing paths would age at different rates (Fig.~\ref{fig:histogram_differences}(middle)). Therefore, the normal distribution (observed at age zero) will morph into a bimodal distribution, where the difference in the mean of two clusters increases over time, highlighting the separation between MAP and LAP group. 
This is the basis for our aged-IC detection. 

However, in a real-world, we have to deal with process drift, reduce the impact of process variation, identify MAP and LAP groups ahead of time, deal with the inaccuracy of tester, account for inaccuracy of the STA and the fact that it does not match the CFST test result. In the next section, we describe our proposed model that deals with each of these phenomenons, for building a reliable aging detection solution. \vspace{-2mm}

\subsection{Proposed Model}\label{proposed_model_section}

Our approach consists of six main steps, described next:
\begin{itemize}
    \item \textbf{(1)} \textbf{ADP set identification:} Selecting a viable set of Age Distinguishing Paths ($ADP$), and dividing it into $MAP$ and $LAP$ subsets.  
    \item \textbf{(2)} \textbf{Building a Golden Timing Model:} Generating a Golden Timing Model (aided by a machine learning regression model) that accounts for process drift, timing prediction of which matches that of CFST test on $MAP$ subset of timing paths. This step intends to model the impact of process drift and systematic process variation.
    \item \textbf{(3)} \textbf{Computing Added Delays:} Inferring the slack of timing paths in the $ADP$ set for both $LAP$ and $MAP$ subsets from the GTM (created in the previous step) as expected value, and from CFST as actual value, and computing the $AD(p)=CFST(p)-GTM(p)$ for each path in each subset. 
    \item \textbf{(4)} \textbf{Inferring MAP-LAP mean shift:} Computing the mean-shift of $AD$ for $MAP$ and $LAP$ subsets; This step intends to reduce the impact of process variation and tester discrete step size on our detection threshold. 
    \item \textbf{(5)} \textbf{Classification:} Using a binary classifier to mark the IC as aged or new. 
\end{itemize}

\subsubsection{\textbf{ADP set identification}}\label{subsubsec:classify} 
Age distinguishing paths consist of two subsets of MAP and LAP timing paths. In order to collect the ADP set and assign timing paths to each of MAP and LAP, we propose the following set of modeling steps, each of which is described next:
\begin{itemize}
     \item Train a regression model for gate-specific age perdition
     \item Build an Aging-Induced Path-Delay Prediction Model
     \item Build an ADP set Classifier
\end{itemize}

\textit{\underline{Regression model for gate-specific age prediction}}: 
To build a model that predicts aging-induced delay increase for each gate, we first create a database that includes each gate type and its corresponding aging-induced delay increase after $i$ months when the gate is fed with different workloads. To emulate different workloads, the gate is simulated under different conditions where in each condition of $COND_{i,j}$, its output signal Duty Cycle and Toggle Count are $DC_i$ and $TC_j$, respectively. Here, Duty Cycle ($DC$) denotes the percentage of the time that the signal is `0'. 

In practice, we consider $I$ different $DC$ and $J$ different $TC$ values for the gate output, and for each condition (among all $I \times J$ conditions) we generate a table of input patterns (among many possibilities) that satisfies the considered toggle count and duty cycle on the output.

To tailor a more precise timing model for each circuit, we determine an approximate range of $TC$s that the circuit's gates outputs experience during run time by simulating the main circuit with a set of random inputs. We use the SAIF file which is generated via simulation to extract the maximum and minimum $TC$s of all signals in the circuit, refer to as $TC_{min}$ and $TC_{max}$ hereafter. Then, we sweep $DC$ in range of $[0,1]$ with the steps of $st$, and $TC$ in range of $[TC_{min},TC_{max}]$ with the steps of $tc$. In this study, 
st and tc are considered as 0.05 and $\frac{TC_{max}-TC_{min}}{50}$, respectively. 

\begin{figure}[t]
\centering
  \includegraphics[width=\columnwidth]{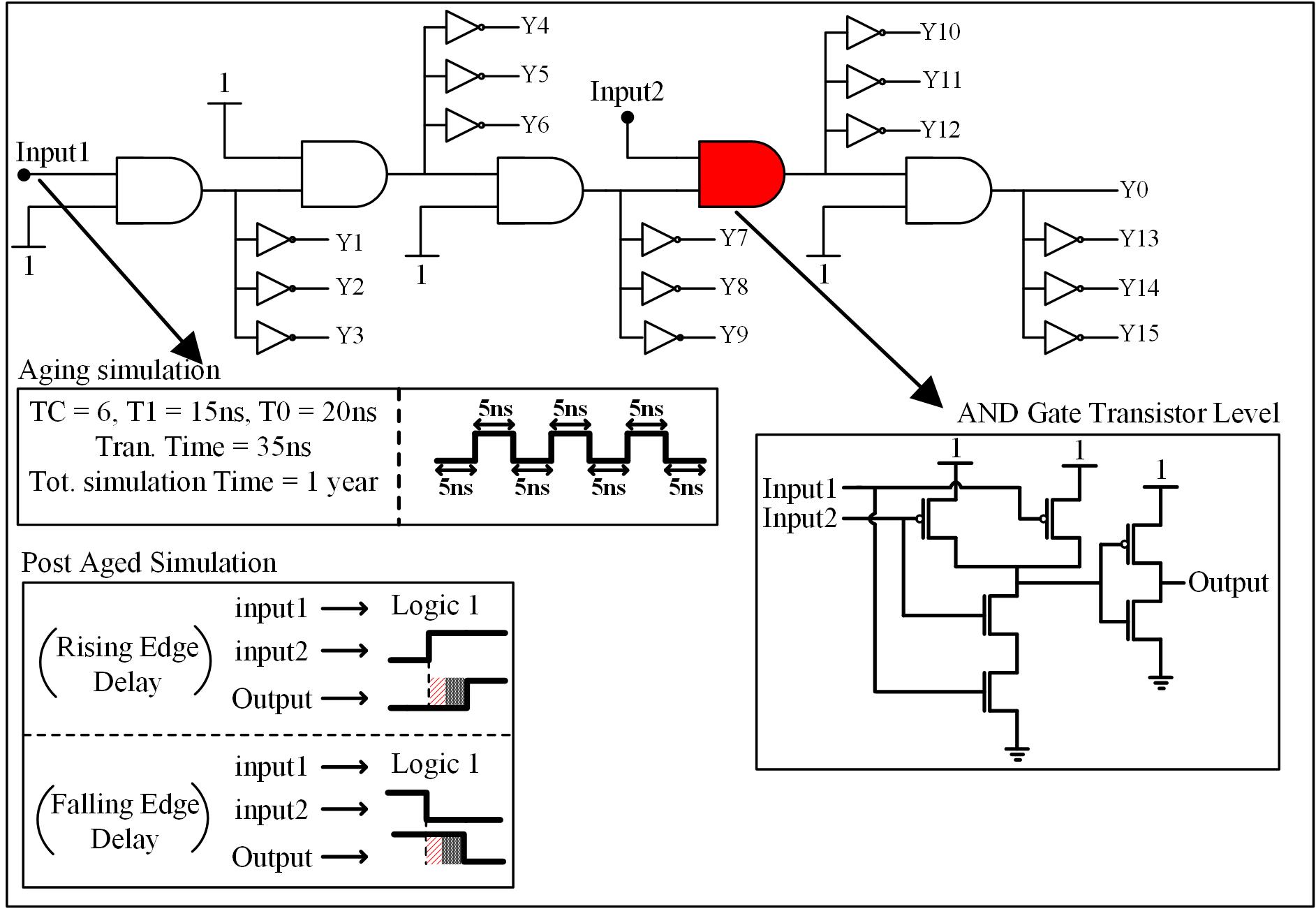}
   \vspace{-1em}
  \caption{SPICE netlist for aging each gate type. \vspace{-4mm}}
 \label{spice_each_gate}
\end{figure}

As the next step, the aging-induced delay change for each gate type in the considered $TC$ and $DC$ combinations are extracted via HSpice aging simulations for $i$ months using the FO4 model for each gate (as shown in Fig.~\ref{spice_each_gate}, and are included in the database which is further deployed for training a \textit{non-linear regression model} to predict the delay of each gate type experiencing unseen $TC$ and $DC$ combinations in its output.

\textit{\underline{Aging-Induced Path-Delay Prediction Model}}: Having the SAIF file used to specify the $TC_{min}$ and $TC_{max}$ and the regression model from the previous step, enables us to infer the aging-induced delay change of each gate in the design after $i$ months of aging, when the device experiences the switching activity and duration cycle in that SAIF file.
Instead of building a new timing model, we use a trick to use existing STA engines for aging-induced path-delay prediction. For this purpose, we replace the delay of each cell, with the delay increment suggested by the regression model (developed in the previous step) and re-time the design. The timing report, in the result of query to the STA tool, provides us with the net delay increase of each timing path, accounting for possible skew in the launch and capture portion of each timing path (if they age differently or have different topology). At this point, we have all information needed to extract the ADP set, described next. 

\textit{\underline{ADP set Classifier}}:  The ADP set, is composed of MAP and LAP subset. Timing paths assigned to each subset have to satisfy two requirements. The first requirement that is common for both MAP and LAP groups is that the selected timing paths should have available slack $s$ in the original (design time) timing model. Where $s$ satisfies the inequality  $f_{max} > 1 /(T-s)$, in which $T$ is clock period, and $f_{max}$ is the maximum clock frequency of the tester. The reason for this path selection is to be able to use CFST to measure the delay of the timing paths, which is needed as a part of our model building. The second requirement is to have a high value of aging-induced path delay prediction for paths in the MAP subset, and low values for timing paths in the LAP subset. 

Identifying MAP and LAP subset could be easily achieved by plotting the histogram of aging-induced path delay predictions for all timing paths that meet the first condition ($f_{max} > 1 /(T-s)$) and identifying set of timing paths that are (predicted to be) least and most affected by aging. This process could be automated by fitting a bimodal function on the resulting plot. 
\begin{equation}
    f = Gauss(\mu_{LAP},\sigma_{LAP}) + Gauss(\mu_{MAP},\sigma_{MAP})
\end{equation}

To differentiate MAP and LAPs from each other, we argue that a timing-path belongs to the MAP group if its aging-induced path delay prediction is greater than $\mu_{MAP} - 2\times\sigma_{MAP}$; while it is included in the LAP group if it is less than $\mu_{LAP} + 2\times\sigma_{LAP}$. Note that the timing paths with mid-range value for aging-induced path delay prediction (that extend the tail of MAP and LAP towards one another) are removed, and are mot included in ADP. 

\subsubsection{\textbf{Building a Golden Timing Model}}\label{subsubsec:GTM} 
The mean of aging-induced delay in LAP and MAP paths deviates from each other as the device ages. We utilize this observation to identify recycled chips. More precisely, the expected delay of timing path, as reported by STA, could be compared with the delay obtained from CFST: $AD=CFST(p)-STA(p)$. The problem with this approach is that the STA timing information could be very off. This might be due to process drift or systematic process variation. In the section below, we describe how each of these fabrication process related phenomenon affects our STA accuracy. Then we describe our proposed learning solution that (by using a set of input features collected from EDA tool, STA engine, and CFST tester on a subset of ADP set) predicts the impact of process drift and systematic process variation. Using this information, and the original STA model, we build a process-drift aware, and systematic-process-variation aware Golden Timing Model (GTM) which will be used in our proposed recycled-IC identification test.   

\textbf{Challenge 1: Process Drift:} 
The Spice model for a new technology node is developed and made available to design houses as soon as the technology is feasible. The spice and technology files are then used to characterize the standard cell libraries, and extract parasitic for physical design. However, the foundry keeps maturing the process over time. Hence, as the process matures, it drifts apart from its snapshot being used by design houses. Since, it is unreasonable to keep updating the technology, spice, and standard cell libraries during physical design, the process matures/drifts while assuring backward compatibility. The most obvious example of process drift is tightening the extent of process variation, a systematic shift in the performance of all-transistor devices, or optical shrinking. 

The process drift causes a \textit{systematic, yet nonlinear} change in the timing behavior of fabricated IC compared to that expected from STA (produced using older process snapshot) and generation of new unused timing slacks. This creates a non-uniform additional margin for different timing paths and improves the yield. However, it makes the comparison of the delays reported by CFST (at test time) and STA (at design time) more difficult. In this case, even an aged IC may have slack more than that reported by STA. Moreover, the added slack is \textit{non-uniformly} distributed, depending on the topology of each timing path, as the process drift affects each timing path differently. Therefore, the STA can not be treated as a fairly accurate model, and can not be used for the detection of aged ICs in our methodology. To remedy this issue, in our proposed solution, we build a learning model (using a subset of ADP paths, features extracted from the EDA tool, and delay reported by CFST) to predict the path-based impact of process drift (for each timing path). Our approach for building the learning model is described below.

\textbf{Challenge 2: Systematic Process Variation:} Systematic process variation is the result of imperfection in one or several process steps, as the result of which, a systematic shift in the behavior of transistors or wires is observed. For example, such a systematic shift may speed up all NMOS transistors, increase the capacitance of a certain metal layer, or reduce the strength of PMOS transistors. Unlike random process variation (mitigation of which will be discussed when we describe our IC classification methodology), the systematic (inter-die) process variation affects all devices similarly. Therefore, systematic process variation behaves very similar to process drift, with the difference that \textit{process drift is the intended consequence of improving the fabrication process}, and the \textit{systematic process variation is an unintended consequence of imperfection in one or several processing steps} (for example if during the chemical mechanical polishing (CMP) step, the height of a certain wire level, e.g. M4, was less or more than the process defined height). The good news is that the systematic process drift can be treated similarly to process drift. In the following text, we describe how we developed a learning model that predicts the impact of both process drift and systematic process variation for ADP set of timing paths.  

\begin{table*}[t]
\centering
\caption{Description for each of 38 features of a sequential circuit, extracted from each timing-path for building the NN dataset. (LP: Launch portion of timing-path, CP: Capture portion of timing-path, DP: Data portion of timing-path, M: Metal Layer, x: drive strength of the gate) \vspace{-4mm}}
\label{feature_table}
\footnotesize
\centering
\scalebox{0.96}{
\begin{tabular}{| l | l | l | l | l | l | l |}
\hline
\multicolumn{6}{|c|}{\textbf{Features}} \\
\hline
Length of M1 in CP 	&  Length of M3 in DP	&   \# cells in DP 		        &    \# cells of x0 strength in  CP   	&   \# cells of x2 strength in  CP	   &     \# cells of x8 strength in  DP     \\     
\hline
Length of M1 in DP 	&  Length of M3 in LP	&   \# cells in LP 		        &    \# cells of x0 strength in  DP	    &   \# cells of x2 strength in  DP	   &     \# cells of x8 strength in  LP     \\                    
\hline
Length of M1 in LP 	&  Length of M4 in CP   &   Setup time		            &    \# cells of x0 strength in  LP	    &   \# cells of x2 strength in  LP 	   &     \# Total fanout                    \\
\hline
Length of M2 in CP 	&  Length of M4 in DP  	&   Delay of CP			        &    \# cells of x16 strength in DP	    &   \# cells of x4 strength in  CP     &      --                           \\
\hline
Length of M2 in DP 	&  Length of M4 in LP  	&   Delay of DP	                &    \# cells of x1 strength in  CP	    &   \# cells of x4 strength in  DP     &      --                           \\ 
\hline
Length of M2 in LP 	&  Length of M5 in DP	&   Delay of LP	                &    \# cells of x1 strength in  DP	    &   \# cells of x4 strength in  LP     &      --                           \\ 
\hline 
Length of M3 in CP 	&  \# cells in  CP      &  Path delay reported in STA   &    \# cells of x1 strength in  LP     &   \# cells of x8 strength in  CP     &      --                           \\ 
\hline
\end{tabular}						
}
\normalsize
\end{table*}

\underline{\textbf{Solution: Learning Model:}} To predict the timing impact of process drift and systematic process variation, we deploy a learning model that is trained to predict the difference between the STA reported delay (at design time) and CFST reported delay (at test time). For this purpose, we first create a dataset (for training, validation, and test) where each instance is a timing-path in the design. Each sample (timing path) is characterized using 38 features. These features are extracted from the GDSII file and the STA engine. Moreover, each sample is given label $L=AD$, where $L=CFST(p)-STA(p)$. Table \ref{feature_table} shows the features considered for a sequential circuit.

To build an accurate learning model, we deployed and trained a stacking regression model~\cite{breiman1996stacked}, aka as stacked generalization \cite{wolpert1992stacked}. The samples used for training are timing paths in the MAP subset of the ADP set. The selected samples are divided into (60\%, 20\%, 20\%) subset to form training, validation, and test set accordingly. The stacking learning model is a two-level ensemble of regressors. Each regressor by itself can be used as a standalone prediction model; however, by stacking them, a more generalized model is obtained that outperforms each of the individual regressors.

\begin{figure}[b]
\centering
  \includegraphics[width=\columnwidth]{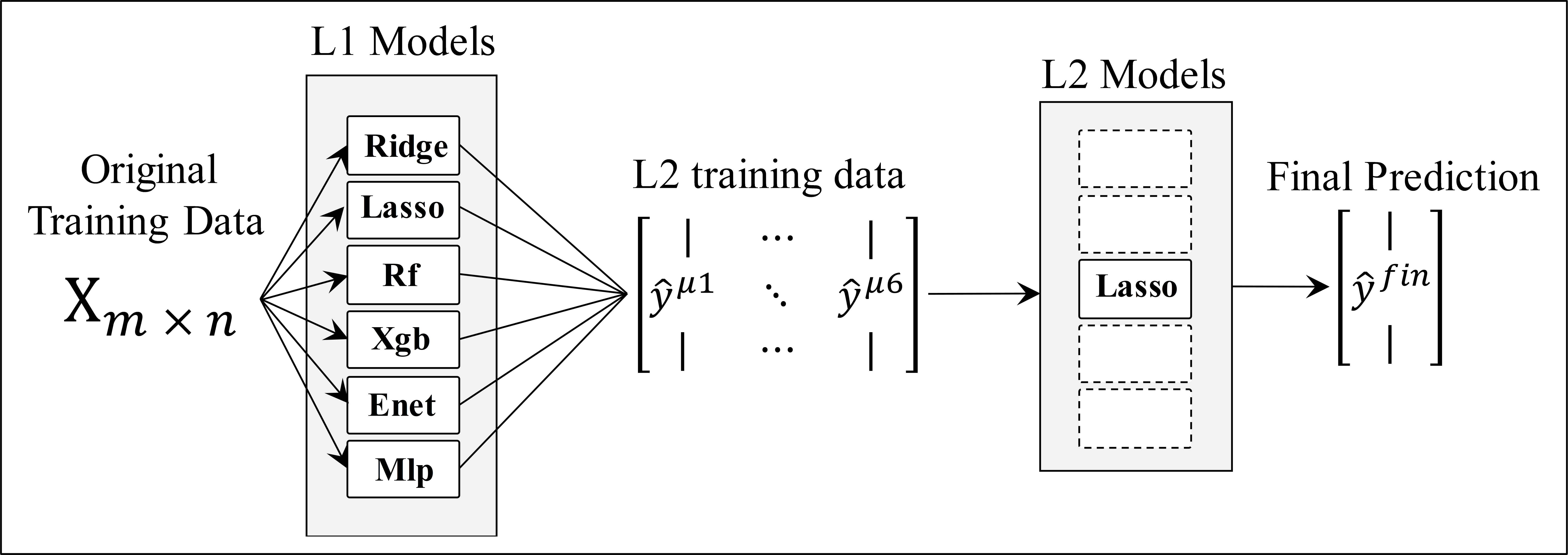}
  \caption{A general view of the used  2-layer stacked model.}
 \label{regressor}
\end{figure}

An abstract view of a two-layer stacked regressor is shown in Fig. \ref{regressor}, in which the first layer, $L_1$, contains of six regressors Xgb\cite{chen2016xgboost}, Enet\cite{zou2005regularization}, Lasso\cite{tibshirani1996regression}, Ridge\cite{gruber2017improving}, MLP\cite{hastie2009elements} and RandomForest\cite{ho1995random}. The second layer can have several regressors, yet we only use Lasso for this layer. More precisely, the pre-processed train-set that contains $m$ instances and $n$ features ($X_{m \times n}$ in Fig. \ref{regressor}) is fed to each of the embedded regressors at the level one, and their outputs, $\hat{y}^{\mu_1}$ to $\hat{y}^{\mu_6}$ in Fig. \ref{regressor}, are fed to the $L_2$ model. The output of Lasso in $L_2$, which is the only regressor in the second layer, is the final prediction, and is specified with $\hat{y}^{fin}$.

We employed Sklearn \cite{sklearn_api} framework to train and test the stacked model. We then used the specified hyper-parameters in Table \ref{hyper} to tune each model. The detail of the implementation can be found at \cite{models_}. Alg.\ref{aging_indentify} summarizes the steps taken in this research for training the NN aiming at identifying recycled chips.

\begin{table}[h]
    \caption{hyper-parameters of each one of the stacked models showed at the Fig. \ref{regressor}. \vspace{-4mm} }
    \footnotesize
    \begin{tabular}{|l|p{7cm}|}
    \hline 
    Model & Hyper-Parameters\tabularnewline
    \hline 
    \hline 
    Ridge & alpha=1, max\_iter=5000\tabularnewline
    \hline 
    Lasso & alpha=0.001, max\_iter=5000\tabularnewline
    \hline 
    RF & n\_estimators=1024, bootstrap=True, min\_leaf=1, min\_split=2\tabularnewline
    \hline 
    Xgb & n\_estimators=1024, learning\_rate=0.05\tabularnewline
    \hline 
    Enet & alpha=0.001, max\_iter=1000\tabularnewline
    \hline 
    {\begin{tabular}[c]{@{}c@{}} \\ MLP \end{tabular}} & in\_layer=42, hidden\_layer=23, out\_layer=1, activation='tanh', optimizer='adam',
    learning\_rate='adaptive', start\_lr='0.1'\tabularnewline
    \hline 
    \end{tabular}
    \normalsize
    \label{hyper}

\end{table}

Let's assume that $Model(p)$ is the prediction of the stacked learning model for the added delay of timing path $p$. Our Golden Timing Model is defined as:
\begin{equation}
    GTM(p) = STA(p) + Model(p)
\end{equation}

\begin{algorithm}[b]
\caption{Training NN }\label{aging_indentify}
\begin{algorithmic}[1]
\scriptsize

\ForAll{\textcolor{black}{$Paths$} in \textcolor{black}{$Benchmark$}} 
    \State \textcolor{black}{$features$ $\gets$ $Collect \ features \ from$ GDSII}
    \State \textcolor{black}{$labels$ $\gets$ $Collect \ Slack \ from \ $ CFST of Fabricated chip}
\EndFor
\State \textcolor{black}{$dataset$} $\gets$ $[features, labels]$
\State \textcolor{black}{$dataset_{MAP}$} $\gets$ $MAP \ of \ dataset$
\State \textcolor{black}{$NN_{model}$} $\gets$ Train a NN with $dataset_{MAP}$
\normalsize
\end{algorithmic}
\end{algorithm}

\begin{figure*}[t]
\centering
  \includegraphics[width=2\columnwidth]{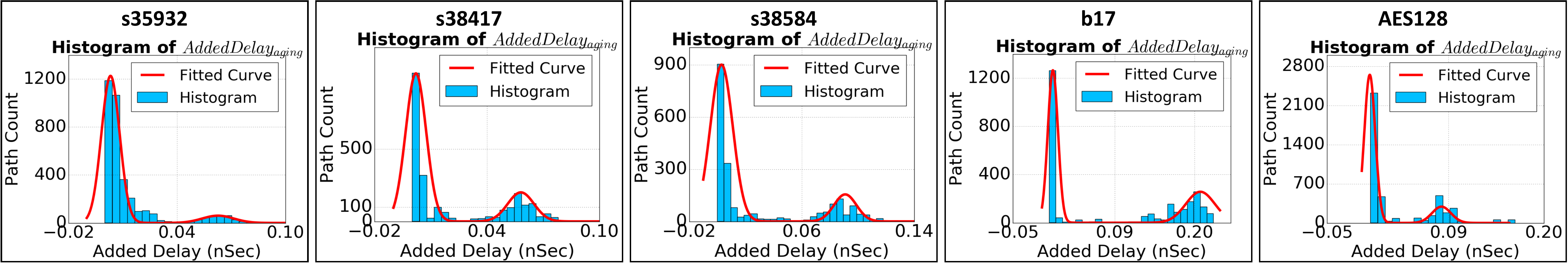}
  \caption{Histograms depicting delay-increase on timing-paths used for classification after one month of aging. For each benchmarks, there exists a bimodal distribution for the $AD$ distinguishing the MAP and LAP paths from each other. \vspace{-4mm}}
 \label{histogram_results}
\end{figure*}

\subsubsection{\textbf{Computing Added Delays}}\label{sec:added-delay}
Now that we have the GTM for the ADP set, we can compute the Added Delay (AD) of each path from: 
\begin{equation}
    AD(p) = CFST(p) - GTM(p)
\end{equation}

Note that this model is tuned to predict the delay of the MAP subset of ADP for the current chip, that could be aged or new. If the IC is new, the $AD$ value for MAP and LAP subset should fall in the same distribution (when collected across many timing paths). However, if the IC is aged, the $Model(p)$ prediction (which is tuned for MAP subset) will be incorrect, resulting in large inconsistency between $CFSF(p)$ and $GTM(p)$. 

\subsubsection{\textbf{Inferring MAP-LAP mean shift}}\label{sec:map-lap}
 Assuming there are $n$ paths in MAP subset, and $m$ paths in LAP subset, we define the mean-shift $MS$ as the 

\begin{equation}\label{eq:mwan-shift}
    MS = \frac{1}{n}\sum_{p \in MAP}{AD(p)} - \frac{1}{m}\sum_{p \in LAP}{AD(P)}
\end{equation}

\subsubsection{\textbf{Classification}}
The last step for detecting aged ICs is a classification based on a simple thresholding mechanism, Using threshold value $Th$, the IC is identified as aged when $Th \le MS$. Choosing a value for $Th$ introduces a trade-off between false positive and sensitivity of the test. The smaller $Th$ the value, the more sensitive the test, and could even identify slightly aged devices at the expense of possible higher false-positive rate. In this paper, we set the threshold to the step size of the CFST tester (to reduce the false positive rate), which we assumed to be 10ps. However, note that there are other mechanisms that could be justified for setting the threshold such as 1) goal-driven threshold to identify devices aged more than $m$ months, 2) error-driven thresholds (such as $m\sigma$ of the error of neural network), 3) simulation-driven thresholds based on the average change of delay of affected timing paths after $m$ months of aging, etc. each of which could be justified based on the ICs use case.

\textbf{A Note on the impact of Random Process Variation:} 
Process variation is variation in the electrical and physical property of transistor devices due to the physical limitations of the fabrication process at scaled geometries \cite{4479829, isqed}. In higher abstract level, it could be modeled as zero-mean variation in threshold voltage of transistor device, such that two identical transistors may end up with different drive strength after fabrication. In this section, we explain why process variation does not impact our aged-IC detection solution.

When considering the impact of process variation in delay of timing paths, the impact of variation in individual devices accumulates and results in the variation in the delay of timing paths. For modeling purpose, we can denote the variation in the delay of timing path $p_i$ using random variable $X_i$, where $E(X_i) = \mu = 0$, and V=$VAR(X_i) = \sigma^2$. Our detection methodology relies on identifying the mean shift between the LAP and the MAP group. Let's assume the MAP group.  The mean of added delay (AD) for timing paths in MAP can be denoted by random variable $\Bar{X}$, where

\begin{equation}
\footnotesize
    \Bar{X}=\frac{1}{n}\sum_{p_i \in MAP}{X_i}
\end{equation}

With this information:
\begin{equation}
\footnotesize
    E(\Bar{X}) = E(\frac{1}{n}\sum_{p_i \in MAP}{X_i}) = \frac{1}{n}E(\sum_{p_i \in MAP}{X_i}) = 0
\end{equation}
\begin{equation}
\footnotesize
    VAR(\Bar{X}) = VAR(\frac{1}{n}\sum{X_i}) = \frac{1}{n^2}VAR(\sum{X_i}) = \frac{n\sigma^2}{n^2} = \frac{\sigma^2}{n}
\end{equation}

The same analogy applies to the mean of AD computed for the LAP group. In another word, the mean shift (used for detection) is a 0-mean random variable with standard deviation $\sigma/\sqrt{n}$, where n represents the number of paths in MAP or LAP group. Therefore, by choosing a large number of paths (n) for each of LAP and MAP set (we collect thousands), the impact of process variation on mean-shift value becomes negligible, and PV does not affect the final classification.

\section{Experimental Results and Discussions}\label{sec:Experim} 
We targeted 5 different IPs including s35932, s38417, s38584, b17, and AES128 from IWLS benchmark suite\cite{iwls2} and hardened them using a commercial 32nm technology via the Synopsys EDA toolset \cite{synopsys}. 
We used Synopsys HSpice for the transistor-level simulations, and the HSpice built-in MOSRA Level 3 model to assess the effect of NBTI and HCI aging~\cite{mosra}. The aging simulations were performed under temperature= 125\textdegree C and Vdd=0.85V for 12 months with 1-~month steps.

For each benchmark, using the Synopsys PrimeTime tool, we extracted N=10 longest paths feeding each endpoint (flip-flop or primary output). To account for the tester frequency step size, in our experiments we only select a subset of these paths whose delay is at least 250ps, resulting in the selection of 3455, 2390, 2121, 2626, and 21460 timing paths for the s35932, s38417, s38584, b17, and AES128 benchmarks, respectively.

To take the impact of process variations into account, in our simulation-based setup, the random patterns we use to generate our GTM is different from the set of patterns we use in aging simulations to extract the aging-induced path delays and creating the ADP classifier. Note that the ADP set classifier is unique for each GDSII netlist and is generated per design. Using the ADP set classifier, we fit a bimodal curve on ADP set's histogram for each design to identify MAP and LAP groups. The histograms and fitted bimodal curves for each target circuitry after one month of usage are shown in Fig.~\ref{histogram_results}. This figure clearly depicts the deviation of MAP and LAP paths from each other when the device is aged. This observation confirms the applicability of the proposed path classification scheme in detecting recycled chips.

\def\phanm{\llap{$-$}}
\tabcolsep=\dimexpr\tabcolsep+0.31ex\relax
\begin{table*}[t]
\caption{The mean error of each ADP set group for all benchmarks. \vspace{-4mm}}
\label{table:results}
\footnotesize
\scalebox{0.99}{
\begin{tabular}{|c|l|l|l|l|l|l|l|l|l|l|l|l|l|l|}
\hline
\multicolumn{1}{|l|}{\textbf{Benchmark}} & \textbf{Aging (Months)} & \textbf{0} & \textbf{1} & \textbf{2} & \textbf{3} & \textbf{4} & \textbf{5} & \textbf{6} & \textbf{7} & \textbf{8} & \textbf{9} & \textbf{10} & \textbf{11} & \textbf{12} \\ \hline \hline
\multirow{4}{*}{\textbf{s35932}} & \textbf{Train and Test (MAP)} & 0.00 & 0.12 & 0.28 & 0.03 & 0.05 & 0.07 & 0.09 & 0.20 & 0.39 & 0.51 & 0.10 & 0.37 & 0.43 \\ \cline{2-15} 
 & \textbf{Evaluate (LAP)} & 0.63 & \phanm12.27 & \phanm19.39 & \phanm22.32 & \phanm23.88 & \phanm23.87 & \phanm27.47 & \phanm28.54 & \phanm30.72 & \phanm30.91 & \phanm32.15 & \phanm35.71 & \phanm39.04 \\ \cline{2-15} 
 & \textbf{Mean Shift} & \phanm0.63 & 12.39 & 19.67 & 22.35 & 23.92 & 23.94 & 27.56 & 28.74 & 31.11 & 31.42 & 32.25 & 36.08 & 39.47 \\ \cline{2-15} 
 & \textbf{Correctly Identified} & \checkmark & \checkmark & \checkmark & \checkmark & \checkmark & \checkmark & \checkmark & \checkmark & \checkmark & \checkmark & \checkmark & \checkmark & \checkmark \\ \hline \hline
\multirow{4}{*}{\textbf{s38417}} & \textbf{Train and Test (MAP)} & \phanm0.05 & \phanm0.89 & \phanm0.54 & \phanm0.76 & 2.81 & \phanm0.49 & 0.27 & \phanm0.20 & \phanm1.81 & \phanm0.96 & \phanm1.85 & 0.89 & 1.88 \\ \cline{2-15} 
 & \textbf{Evaluate (LAP)} & 0.57 & \phanm28.93 & \phanm31.07 & \phanm32.76 & \phanm33.05 & \phanm33.55 & \phanm36.54 & \phanm40.91 & \phanm41.01 & \phanm43.60 & \phanm44.52 & \phanm49.99 & \phanm62.87 \\ \cline{2-15} 
 & \textbf{Mean Shift} & \phanm0.62 & 28.04 & 30.53 & 32.00 & 35.87 & 33.06 & 36.82 & 40.71 & 39.20 & 42.64 & 42.67 & 50.89 & 64.75 \\ \cline{2-15} 
 & \textbf{Correctly Identified} & \checkmark & \checkmark & \checkmark & \checkmark & \checkmark & \checkmark & \checkmark & \checkmark & \checkmark & \checkmark & \checkmark & \checkmark & \checkmark \\ \hline \hline
\multirow{4}{*}{\textbf{s38584}} & \textbf{Train and Test (MAP)} & 0.00 & \phanm0.20 & \phanm2.54 & 1.40 & \phanm1.43 & 1.79 & 1.22 & \phanm1.05 & \phanm2.99 & \phanm0.30 & \phanm2.92 & \phanm2.86 & \phanm2.95 \\ \cline{2-15} 
 & \textbf{Evaluate (LAP)} & 0.46 & \phanm26.46 & \phanm31.50 & \phanm34.47 & \phanm39.01 & \phanm41.02 & \phanm48.17 & \phanm46.42 & \phanm49.28 & \phanm49.53 & \phanm53.01 & \phanm55.15 & \phanm57.46 \\ \cline{2-15} 
 & \textbf{Mean Shift} & \phanm0.46 & 26.26 & 28.96 & 35.87 & 37.58 & 42.81 & 49.39 & 45.37 & 46.29 & 49.23 & 50.08 & 52.30 & 54.52 \\ \cline{2-15} 
 & \textbf{Correctly Identified} & \checkmark & \checkmark & \checkmark & \checkmark & \checkmark & \checkmark & \checkmark & \checkmark & \checkmark & \checkmark & \checkmark & \checkmark & \checkmark \\ \hline \hline
\multirow{4}{*}{\textbf{b17}} & \textbf{Train and Test (MAP)} & 0.02 & \phanm2.75 & \phanm2.10 & \phanm5.46 & \phanm1.12 & 0.64 & 0.24 & \phanm1.05 & \phanm0.99 & 0.66 & \phanm0.42 & \phanm4.58 & \phanm4.21 \\ \cline{2-15} 
 & \textbf{Evaluate (LAP)} & 0.14 & \phanm38.22 & \phanm40.65 & \phanm44.99 & \phanm48.21 & \phanm53.94 & \phanm57.18 & \phanm68.94 & \phanm73.40 & \phanm91.61 & \phanm100.83 & \phanm104.66 & \phanm107.85 \\ \cline{2-15} 
 & \textbf{Mean Shift} & \phanm0.12 & 35.47 & 38.55 & 39.53 & 47.09 & 54.58 & 57.42 & 67.89 & 72.41 & 92.27 & 100.42 & 100.08 & 103.64 \\ \cline{2-15} 
 & \textbf{Correctly Identified} & \checkmark & \checkmark & \checkmark & \checkmark & \checkmark & \checkmark & \checkmark & \checkmark & \checkmark & \checkmark & \checkmark & \checkmark & \checkmark \\ \hline \hline
\multirow{4}{*}{\textbf{AES128}} & \textbf{Train and Test (MAP)} & 0.02 & 1.06 & 0.87 & 0.65 & 1.94 & 0.35 & 0.31 & 2.09 & 1.28 & 1.84 & 1.36 & 1.49 & 0.63 \\ \cline{2-15} 
 & \textbf{Evaluate (LAP)} & 0.53 & \phanm35.31 & \phanm39.77 & \phanm42.55 & \phanm47.95 & \phanm44.71 & \phanm46.68 & \phanm45.18 & \phanm52.32 & \phanm51.52 & \phanm55.25 & \phanm62.85 & \phanm84.86 \\ \cline{2-15} 
 & \textbf{Mean Shift} & \phanm0.50 & 36.37 & 40.65 & 43.20 & 49.89 & 45.06 & 46.99 & 47.28 & 53.60 & 53.36 & 56.61 & 64.35 & 85.49 \\ \cline{2-15} 
 & \textbf{Correctly Identified} & \checkmark & \checkmark & \checkmark & \checkmark & \checkmark & \checkmark & \checkmark & \checkmark & \checkmark & \checkmark & \checkmark & \checkmark & \checkmark \\ \hline
\end{tabular}
}
\end{table*}

As the next step after classifying the timing paths, for each benchmark, we extracted the features presented in Table~\ref{feature_table} from its GDSII file, and used them to generate 13 datasets per benchmark related to $i$ months of aging where $0\le i\le 12$. Each dataset includes the extracted features and the slacks collected from one of these 13 aging simulations. We used these datasets to generate a unique GTM for each circuit-under-test. 

We deployed the extracted GTM for each benchmark circuits aged between zero and 12 months, and calculated the MAP-LAP mean shift via equation~\ref{eq:mwan-shift}. The results are shown in Table \ref{table:results}. As depicted, the mean shift between LAP and MAP paths significantly increases for an aged device compared to its fresh (age=0) counterpart. The more the device is aged, the higher the value of the mean shift between its LAPs and MAPs. However, as expected the rate of mean-shift increase is higher initially. This is because the aging effect is high in the first couple of months but the aging-induced threshold voltage tends to saturate for long stress times (refer to Fig.~\ref{fig:stress_recovery}). In particular, as Table \ref{table:results} shows for the s35932 benchmark, the mean shift changes from -0.63ps to 12.39ps (20x increase) after 1 month, while it increases 50\% in the following month compared to its value in month 1. The same trend can be observed in other benchmarks. On average, over all benchmarks, the mean shift increases 29.15, 36.02, 44.19, 53.71, and 72.23 ps after 1, 3, 6, 9, and 12 months of aging. Accordingly, the proposed method can accurately differentiate the new and recycled chips from each other.

\section{Conclusion}\label{sec:Concl}
In this paper, we presented a novel methodology for detecting Aged-ICs. Our detection methodology is based on a side-channel delay analysis. Our model is tolerant of random process variation, systematic process variation, and process drift. Our proposed solution does not rely on the existence of a Golden IC. Instead, using features collected from the design, and delay samples collected from a subset of timing paths using the Clock Frequency Sweeping test (CFST), builds a learning model that predicts the delay difference between the CFST and STA. We also presented a methodology to distinguish between two sets of timing paths that age more/less overtime. Using the collected delay information, we compute the delay difference between these two sets of paths (Aged Distinguishing Paths) as a measure of the age of the IC. 

\section{ACKNOWLEDGEMENT}
This work was partly funded by the START program at UMBC and the National Science Foundation CAREER Award (NSF CNS-1943224), and by the Defense Advanced Research Projects Agency (DARPA-AFRL Award \#: FA8650-18-1-7820), and National Science Foundation (NSF Award \#: 1718434) at George Mason University.  

\bibliographystyle{ACM-Reference-Format}

\bibliography{Bibliography.bib}

\end{document}